**Unconventional Magnetization of $Fe_3O_4$ Thin Film Grown on Amorphous $SiO_2$ Substrate**

Zhi-Guo Liu, Shang-Fei Wu, Jia-Xin Yin*, Wen-Hong Wang, Wan-Dong Kong, Hao-Jun Yang, Pierre Richard, Hong Ding, Lei Yan*

Institute of Physics, Chinese Academy of Sciences, Beijing 100190, China

Collaborative Innovation Center of Quantum Matter, Beijing 100190, China

*E-mail: jiaxinyin@iphy.ac.cn; lyan@iphy.ac.cn.

**High quality single crystal $Fe_3O_4$ thin films with (111) orientation had been prepared on amorphous $SiO_2$ substrate by pulsed laser deposition. The magnetization properties of the films are found to be highly unconventional. The Verwey transition temperature derived from the magnetization jump is around 140K, which is higher than the bulk value and it can be slightly suppressed by out-plane magnetic field; the out-of-plane magnetization, which is unexpectedly higher than the in-plane value, is also significantly increased as compared with the bulk value. Our findings suggest that the local Coulomb correlation U and the effective ferromagnetic exchange interaction J of Fe 3d electrons are both dramatically strengthened and out-of-plane directionally entangled by the unusual coupling between $Fe_3O_4$ thin film and the amorphous $SiO_2$ substrate.**

Magnetite ($Fe_3O_4$) is one of the well-known ferrimagnetic materials, which has been attracting a lot of attention due to its unique electrical and magnetic properties, such as low electrical resistivity at room temperature, high Curie temperature (858 K), 100% spin polarization [1-4]. These properties make $Fe_3O_4$ a promising candidate for spintronic devices [5,6].

Magnetite has a cubic inverse spinel structure that is based on a face-centered cubic (fcc) lattice of oxygen anions. The cubic unit cell of $Fe_3O_4$ contains 32 oxygen anions and 24 iron cations, *i.e.* 8 $Fe^{2+}$ and 16 $Fe^{3+}$ ions which occupy interstices of oxygen ions; 8 tetrahedral (A) sites solely occupied by $Fe^{3+}$ ions whereas 16 octahedral (B) sites equally shared by $Fe^{3+}$ and $Fe^{2+}$ ions. The magnetization of magnetite can be viewed as that the magnetic moments within the A and the B sublattices are ferromagnetically aligned while the two sublattices are antiferromagnetic with respect to each other.

At the room temperature, electrons continuously rapidly hop between $Fe^{2+}$ and $Fe^{3+}$ cations at B sites, leading to a fairly low electrical resistivity. Upon cooling, bulk $Fe_3O_4$ undergoes a metal–insulator transition termed as the Verwey transition [1,4]. Below 120 K, the hopping action is frozen and consequently the resistivity is increased by two orders of magnitude with a concomitant decrease in the magnetic moment. This transition is generally from a disordered phase to a charge ordered phase of $Fe^{2+}$ and $Fe^{3+}$ cations [7-11].

Epitaxial $Fe_3O_4$ thin films have some physical properties that differ significantly from those of bulk single crystals, such as distinct transition temperatures on different substrates [12,13]. $Fe_3O_4$ thin films grown on single crystalline substrates naturally form antiphase boundaries (APBs) [14,15], which always produce anomalous properties, for instance, a lower Verwey transition temperature ($T_V$), and the out-of-plane magnetization is smaller than the in-plane magnetization at the same magnetic field, while thin films grown on buffer layers can result in polycrystalline or amorphous phases.

Thermal growth $SiO_2$ substrates are typical amorphous substrates, which do not provide preferred orientation for films grown on them. Hence, the films grow relaxed, leading to its lattice parameter very close to the bulk value [16]. In this paper, a single crystal $Fe_3O_4$ thin film with the (111) orientation on a 300nm thermal growth $SiO_2$ substrate is prepared. Experimental results show that $T_V$ increases to 140 K and can be suppressed by out-plane magnetic field; moreover, the out-plane magnetization is larger than the bulk value and the in-plane magnetization.

The pulsed laser deposition technique (PLD) is an effective method to produce high quality thin films due to the high kinetic energy of atoms and ionized species in the laser-induced plasma. The thin film used was grown on an amorphous $SiO_2$ substrate using an α-$Fe_2O_3$ target, and a 300 nm amorphous $SiO_2$ substrate is grown on a Si (100) substrate. The target used for the ablation has been prepared by a standard solid-state reaction method. A KrF excimer laser source (λ=248 nm, pulse width=20 ns) was used to ablate the target at a pulse repetition rate of 10 Hz, and a fluence of 250 mJ pulses$^{-1}$ was directed at a 45 °angle of incidence on the target. The distance from the target to the substrate was maintained at 60 mm during deposition. Before the deposition, the chamber was evacuated to a pressure of $9.9 \times 10^{-6}$ Pa and the deposition was carried out at a substrate temperature of 575 ℃ and in a vacuum of $6\times10^{-3}$ Pa. During the deposition the target was rotated at a rate of 3 rpm to avoid excessive heating and erosion at a single spot on the target surface. At the end of the deposition, the film was cooled down to the room temperature at 3 ℃ $min^{-1}$ in the same environment as used during the deposition.

The structural of the deposited film was characterized by x-ray diffraction (XRD) in θ-2θ geometry using a Cu Kα radiation (λ=1.54059 Å) (Rigaku, Japan). Phase purity of the film was checked by performing laser Raman spectroscopy. Raman spectra was recorded using a HR800 Jobin-Yvon spectrometer employing He–Ne laser (λ=632.8 nm). The measured resolution of the spectrometer is 1 $cm^{-1}$. Both XRD and Raman measurements were performed at the room temperature. The film thickness was determined by scanning electron microscope (SEM) and estimated to be 140 nm. The magnetization measurements were carried out using a SQUID vibrational sample magnetometer (SVSM) and the R(T) measurements were performed in standard four-probe geometry using a Quantum Design PPMS.

Figure. 1(a) shows the XRD spectrum of a $Fe_3O_4$ thin film on a $SiO_2$ substrate, which clearly suggests that the film is grown with a preferred orientation in the (111) direction. No impurity or peak from other phase of iron oxide peaks is detected from the XRD patterns, suggesting that the film has a pure (111) orientation. The full width at half maximum for the rocking curve of the (111) peak of $Fe_3O_4$ is 0.1°, indicating a high degree of orientation quality. Analysis of the XRD peaks shows the presence of the $Fe_3O_4$ phase in the films. However, the possibility of the presence of γ-$Fe_2O_3$ cannot be totally ruled out since they have nearly the same lattice parameters (a = 0.8397 nm for $Fe_3O_4$; a = 0.8342 nm for $\gamma$-$Fe_2O_3$; and a = 0.840 nm for the thin film as calculated from XRD) and all diffraction peaks of $Fe_3O_4$ and $\gamma$-$Fe_2O_3$ appear at nearly the same $2\theta$ values.

To confirm the phase purity of $Fe_3O_4$ in the film, Raman spectroscopy measurements have been performed since the Raman technique is very sensitive to the different phases of iron oxides on account of the vibrational frequencies of different compositions [17]. Figure. 1(b) shows a Raman spectrum of a $Fe_3O_4$ thin film on a $SiO_2$ substrate. We can see an acceptable consistency of the $T_{2g}(1)$, $T_{2g}(3)$, $T_{2g}(2)$ and $A_{1g}$ modes corresponding to the $Fe_3O_4$ phase at 193, 306, 540 and 669 $cm^{-1}$, respectively. These values are close to that observed in a magnetite single crystal [18]. We note that none of the Raman spectra reveal any signature of the corresponding modes of $\gamma$-$Fe_2O_3$ normally seen at 350, 500, and 700 $cm^{-1}$. Thus, Raman studies further confirm the pure single phase of our $Fe_3O_4$ thin film.

Thin films can usually be very differect from the bulk materials due to its interaction with the substrate. Here we especially focus on the magnetization properties of the magnetite thin film. Figure. 2 shows the normalized zero-field-cooled (ZFC) and field-cooled (FC) magnetization as a function of temperature with magnetic fields applied parallel and perpendicular to the film plane, respectively. The ZFC spontaneous magnetization curves exhibit a spontaneous magnetization jump in the vicinity of $T_V$= 140 K. Thin films of $Fe_3O_4$ are known to usually exhibit suppressed $T_V$ (<120K) that could originate from a variety of other reasons, like strain or APB. However, the measured results on our films show that the transition temperature of $Fe_3O_4$ is enhanced, which means that there must be an unusual positive coupling between the film and the amorphous $SiO_2$ substrate. More interestingly, $T_V$ associated with in-plane magnetization is unchanged as the applied field increases, while the transition temperature decreases for out-plane magnetization. The Verwey transition is usually viewed as a charge ordering of the $Fe^{2+}$ and $Fe^{3+}$ in the B sublattice due to local Coulomb correlation U [7-11]. Accordingly, the enhancement of this transition may indicate that the U is strengthened by the interaction with the $SiO_2$ substrate; and its coupling with the external out-of-plane magnetic field is also unusual, suggesting the charge and spin degrees of freedom are more strongly entangled on our thin film than in the bulk material.

The magnetization hysteresis loops of our thin film also exhibits unconventional behaviour. Figure. 3 shows ZFC magnetization hysteresis loops with magnetic fields parallel and perpendicular to the film plane at different temperatures. We observe that the magnetization of the $Fe_3O_4$ film is unsaturated at 50 kOe, with a slight residual slope extending to larger magnetic fields. The in-plane hysteresis curves display an almost rectangular shape with higher remanence and smaller coercivity than that of out-of-plane case at the same temperature, indicating that the magnetic moments lie in plane, which is in accordance with the assumption that the easy axis of the films lies in the film plane. The magnetization values measured at 50 kOe field and the coercivities at different temperatures are listed in Table I. At 50 kOe, the out-plane magnetization is about 530 emu cm$^{-3}$ and not saturated even at such a high field at room temperature.

This value is significantly higher than the saturation magnetization of 471 emu cm$^{-3}$ measured at 300 K for bulk Fe$_3$O$_4$. It implies that the film is free from APBs since strong anti-ferromagnetic coupling within APBs reduces the magnetization [19,20]. The enhancement of the saturation magnetization may favour the interpretation that the intrinsic effective ferrimagnetic exchange interaction J is strengthened by the interaction with SiO$_2$ substrate. Here the enhancement of this effective J may be related to the decreasing of the antiferromagnetic exchanged interaction or increasing of the ferromagnetic exchange interaction of Fe 3d electrons. More unexpectedly, through the comparison of the values of magnetizations with 50 kOe field parallel and perpendicular to the film plane, the out-plane magnetization is even higher than the in-plane magnetization. This is very striking as the shape anisotropy is expected to confine the magnetic moments to the plane of the film; and it implies the existence of unconventional ferrimagnetic coupling anisotropy. Recalling that the Verwey transition can lead to a negative jump of the magnetization while the out-plane magnetic field can supress the Verwey transition for our film, it may be inferred that the Coulomb correlation and the ferrimagnetic interaction is entangled along c-axis of the film and contribute to their unusual anisotropy.

In summary, the (111) oriented single crystalline Fe$_3$O$_4$ thin films have been successfully prepared on amorphous SiO$_2$ substrates. The Verwey transition temperature is higher than the 120K transition temperature of the bulk material and can be suppressed by the out-of-plane magnetic field; moreover, the out-of-plane magnetization is higher than that of the bulk material and the in-plane magnetization.

These observations indicate that the local Coulomb correlation and ferrimagnetic exchange interaction are both enhanced and out-of-plane directionally coupled by the unconventional interaction between the $Fe_3O_4$ thin film and the amorphous $SiO_2$ substrate. Such novel magnetization properties not only have potential application on spintronics, but also offer an exciting platform for future theoretical and experimental investigations on the cooperation or competition behaviors of U and J in this correlated system.

**Acknowledgements**

This work was supported by grants from Chinese Academy of Sciences (2010Y1JB6), Ministry of Science and Technology of China (2010CB923000, 2011CBA001000, 2011CBA00102, 2012CB821403 and 2013CB921703) and Chinese National Science Foundation (11234014, 11227903, 11004232, 11034011/A0402 and 11274362).

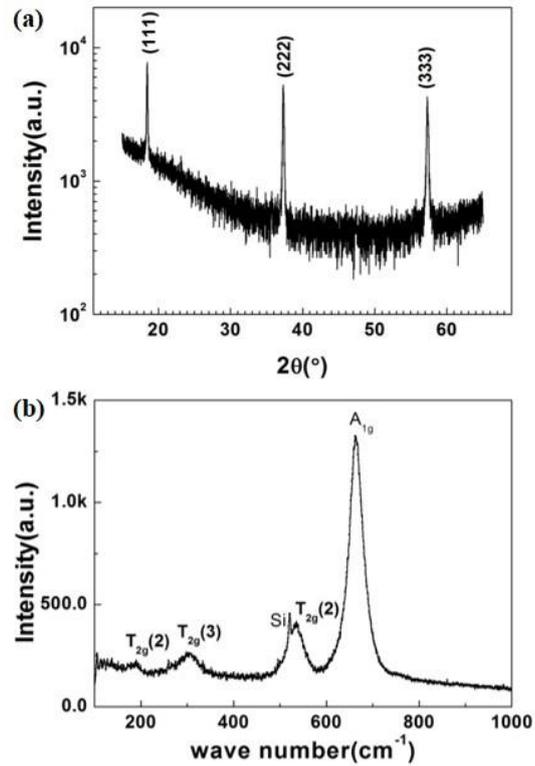

FIG. 1 (color online) (a) XRD spectrum of a $Fe_3O_4$ thin film on a $SiO_2$ substrate measured at the room temperature. (b) Raman spectrum of a $Fe_3O_4$ thin film on a $SiO_2$ substrate measured at the room temperature.

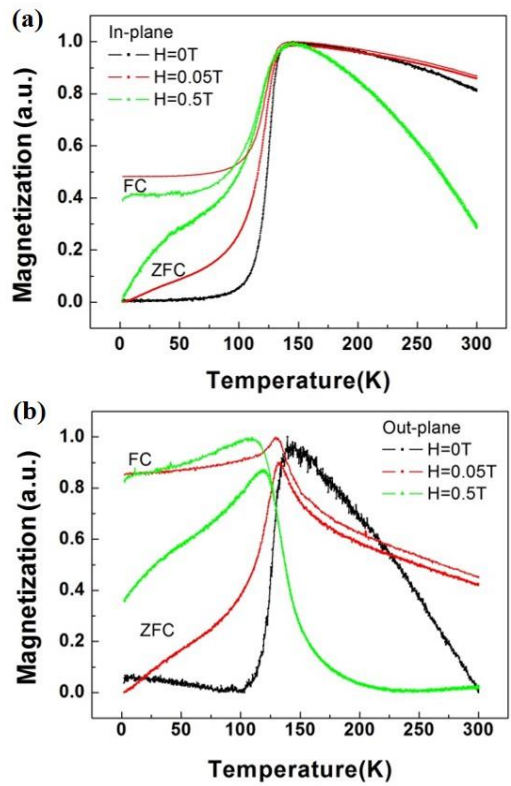

FIG. 2 (color online) Normalized ZFC and FC magnetization as a function of temperature with magnetic fields applied parallel (a) and perpendicular (b) to the thin film plane.

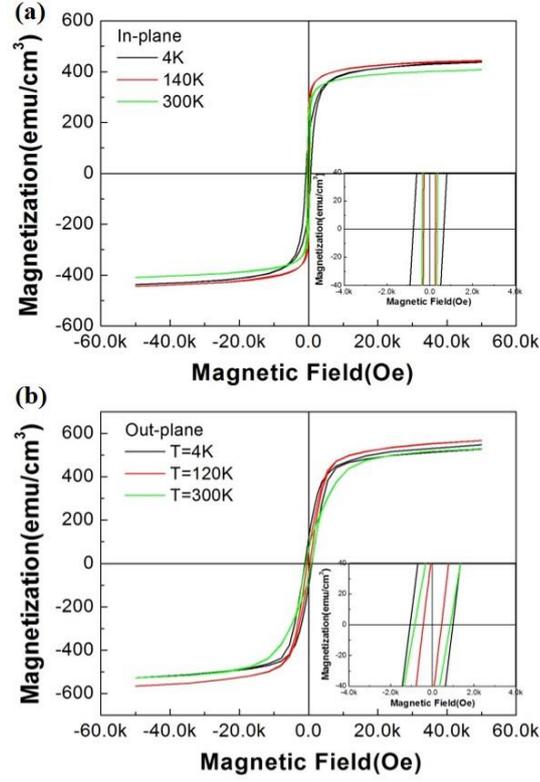

FIG. 3 (color online) Magnetization hysteresis loops measured at 4, 140 and 300 K with an in-plane (a) and out-of-plane (b) magnetic field up to ±50 kOe. The inset shows the low field magnetization.

|  | Field | 4K | 140K | 300K |
| --- | --- | --- | --- | --- |
| $M_{5T}$ (emu cm$^{-3}$) | In-plane | 437 | 442 | 409 |
|  | Out-plane | 528 | 567 | 530 |
| $H_c$ (Oe) | In-plane | -766/664 | -274/250 | -359/359 |
|  | Out-plane | -1071/974 | -445/432 | -847/848 |

TABLE I. Temperature dependence of $M_{5T}$ and $H_c$ of the $Fe_3O_4$ film obtained in the H-parallel and H-perpendicular cases with respect to the film plane.